# CHAPTER 4

## CNIDARIA: fast, reference-free clustering of raw and assembled genome and transcriptome NGS data

**Aflitos, Saulo Alves, Edouard Severing, Gabino Sanchez-Perez, Sander Peters, Hans de Jong, Dick de Ridder. CNIDARIA fast, reference-free clustering of raw and assembled genome and transcriptome NGS data, accepted for publication by *BMC Bioinformatics***

## Summary

**Background**: Identification of biological specimens is a major requirement for a range of applications. Reference-free methods analyse unprocessed sequencing data without relying on prior knowledge, but generally do not scale to arbitrarily large genomes and arbitrarily large phylogenetic distances.
**Results**: We present CNIDARIA, a practical tool for clustering genomic and transcriptomic data with no limitation on genome size or phylogenetic distances. We successfully simultaneously clustered 169 genomic and transcriptomic datasets from 4 kingdoms, achieving 100% identification accuracy at supra-species level and 78% accuracy for species level.
**Discussion**: CNIDARIA allows for fast, resource-efficient comparison and identification of both raw and assembled genome and transcriptome data. This can help answer both fundamental (e.g. in phylogeny, ecological diversity analysis) and practical questions (e.g. sequencing quality control, primer design).

## Background

Unequivocal identification of biological specimens is a major requirement for reliable and reproducible (bio)medical research, control of intellectual property by biological patent holders, regulating the flow of biological specimen across national borders, enforcing the Nagoya protocol (Diversity, 2010) and verifying the authenticity of claims of the biological source of products by customs authority.

Several methods for species identification have been developed based on DNA analysis, that can be classified as probe-based and nucleotide sequencing based methods. Probe-based technologies include microarrays, PCR probes, DNA fingerprinting and immunoassays involving the hybridization of DNA samples with predetermined sets of probes or primers. Such methods are cheap and allow precise identification, but may fail in cases where target DNA is not precisely matched by the probes or primers. Alternatively, nucleotide sequencing methods have been developed to increase accuracy, flexibility and throughput. These can be separated into complete or targeted approaches. Targeted identification of short and highly variable genomic regions by exome capture, Expressed Sequence Tag (EST), DNA barcoding and ribosomal DNA (rDNA) sequencing has been used for many years. Targeted DNA sequencing can be done iteratively for taxonomic identification at subspecies, accession and cultivar levels. Whole Genome Sequencing (WGS) and RNA-seq using Next Generation Sequencing (NGS) technology, examples of complete sequencing methods, have the highest information content of all methods, although its high cost has prevented it from being adopted massively. However, with the recent reduction of costs and increase in throughput, NGS starts to become more prevalent, making it a feasible alternative method for species identification. This calls for the creation of a new set of tools to comprehensively analyse the deluge of data. Methods for species identification based on NGS data can be separated into two main classes: reference-based and reference-free methods (reviewed in Pettengill *et al.*, 2014). Reference-based methods usually map the sequence reads to the genome of a close rel-ative and infer the phylogeny by aligning the observed polymorphisms. This technology requires quality control (cleaning) of the data, mapping the data to the genomic sequence of a close relative, and detection and comparison of polymorphisms (Bertels *et al.*, 2014). In contrast, reference-free methods (RFMs) are designed to analyse unprocessed sequencing data without any previous knowledge of its identity. The data can be compared against other datasets of unknown samples, in the case of metagenomics comparing population structures (Chan *et al.*, 2008a; Chan *et al.*, 2008b; Diaz *et al.*, 2009; Greenblum *et al.*, 2015; Hurwitz *et al.*, 2014; McHardy and Rigoutsos, 2007; Schloissnig *et al.*, 2013; Smits *et al.*, 2014; Wood and Salzberg, 2014; Yang *et al.*, 2010) or against a panel of known species. In the latter case, it can identify previously unknown samples, giving it an approximate position relative to the known species.

RFM methods can be based on the Discrete Fourier Transform (DFT), compression and *k*-mers. DFT methods, such as in (Hoang *et al.*, 2015), transform nucleotide sequences into frequency statistics and compare these for species classification. Although remarkably fast, this method is not able to store the differences between the genomes for further enquiry, yielding no insight into sequence composition. Compression based methods calculate the distance between pairs of sequences by analysing the reduction in computer memory usage when both sequences are compressed together (Tran and Chen, 2014). However, compression-based methods are time and resource intensive for large genomes or large datasets. *K*-mer based methods split the nucleotide sequence information in the form of NGS reads (.fastq files) or assembled data (.fasta files) into all its constituent substrings of size k, which are then used to calculate the similarity between the sequences of the samples. Several implementations of *k*-mer based RFMs exists, such as FFP (Sims *et al.*, 2009), Co-Phylog (Yi and Jin 2013), NextABP (Roychowdhury *et al.*, 2013), MultiAlignFree (Ren *et al.*, 2013), kSNP (Gardner and Hall, 2013) and Spaced Words/KMACS (Horwege *et al.*, 2014). Although their underlying principles are generally useful for the analysis of large data collections, most implementations are designed for either analysis of a limited portion of the data, such as organelles or ribosomal DNA, or analysis of closely related species (such as bacteria, in metagenomics applications). As a consequence, it is not feasible to apply these tools on large amounts of whole-genome sequencing data or to analyse data that spans large phylogenetic distances. One exception is the tool proposed in (Cannon *et al.*, 2010) that is able to find polymorphisms shared by subsets of the data by counting and merging sets of *k*-mers. This tool was successfully used in (Kua *et al.*, 2012) to compare 174 chloroplast genomes and has a similar approach to ours.

Here we present Cnidaria, an algorithm that employs a novel RFM strategy for species identification based on *k*-mer counting, designed from the ground up to allow analysis of very large collections of genome, transcriptome and raw NGS data using minimal resources. Cnidaria improves over previous methods and overcomes their limitations on size and phylogenetic distance by allowing fast analysis of complete NGS data. To this end, it can export a database with pre-processed data so that new samples can be quickly compared against a large database of references, without the need to re-process all the data. In contrast to the method used by REFERENCEFREE (Cannon *et al.,* 2010), CNIDARIA is much faster, produces smaller files, is able to produce phylogenetic trees and uses the popular and fast *k*-mer count software JELLYFISH (Marcais and Kingsford, 2011), allowing for easy integration in existing NGS quality checking pipelines. We demonstrate the performance and capabilities of CNIDARIA by analysing 169 samples, achieving excellent identification accuracy.

## Implementation

CNIDARIA works with both raw sequencing data and assembled data, both from WGS and RNA-seq sources, in any combination. It uses *k*-mers extracted by JELLYFISH (Marcais and Kingsford, 2011), a fast *k*-mer counting tool that produces a database of all *k*-mers present in a query sequence. The advantage of JELLYFISH over comparable software is its ability to create a sparse, compressed database in which the *k*-mers are ordered according to a deterministic hashing algorithm, thus allowing for the parallel and efficient merging/processing of the databases once all *k*-mers are in the same predictable order across different databases.

After creating a database containing all *k*-mers in each sample by running JELLYFISH, CNIDARIA efficiently merges these databases and extracts the number of *k*-mers shared between the samples. CNIDARIA then converts this data into a matrix containing the number of *k*-mers shared between each pair of samples. This information is then used to calculate the *Jaccard* distance, as in CO-PHYLOG (Yi and Jin, 2013), between samples:

$$D_{Jaccard} = 1 - \frac{V_{ab}}{V_a + V_b - V_{ab}}$$

Here, $V_{ab}$ is the number of *k*-mers shared by both samples A and B, Va is the number of valid *k*-mers in sample A and Vb is the number of valid *k*-mers in sample B. When A is equal to B, the distance is 0. The resulting *Jaccard* distance matrix is then processed by PYCOGENT v.1.5.3 (Knight *et al.*, 2007), which clusters the data using Neighbour-Joining and creates a phylogenetic tree in NEWICK format, aiding in the identification of the unknown samples in the dataset. For easy visualization of the data, the summary database can also be converted to a standalone HTML page for (dynamic) display of the phylogenetic tree and plotting any statistics of the analysis directly in the tree. A graphical representation of these steps can be found in Figure 1.

CNIDARIA can be run in two modes: Database Creation Mode and Sample Analysis Mode. The latter is an order of magnitude faster than the former, generating only a CNIDARIA Summary Database (CSD); Database Creation Mode takes longer to run as it exports both a CSD file and a Cnidaria Complete Database (CCD). The CSD contains the total number of *k*-mers for each sample, the number of *k*-mers shared by at least two samples (valid *k*-mers), and the pairwise number of shared *k*-mers. The CCD file contains all *k*-mers present in the datasets analysed, stored in the database using two bits per nucleotide encoding, and their respective presence/absence list describing which set of samples contains each *k*-mer. The CCD can be used as an input to Cnidaria itself, allowing new samples to be directly compared against a pre-calculated larger dataset, speeding up the analysis significantly since the speed of Cnidaria is directly correlated to the number and size of the input files. Hence, the software permits a shorter run time for the comparison of a new sample, using Sample Analysis Mode, against a large reference panel stored in a single CCD file.

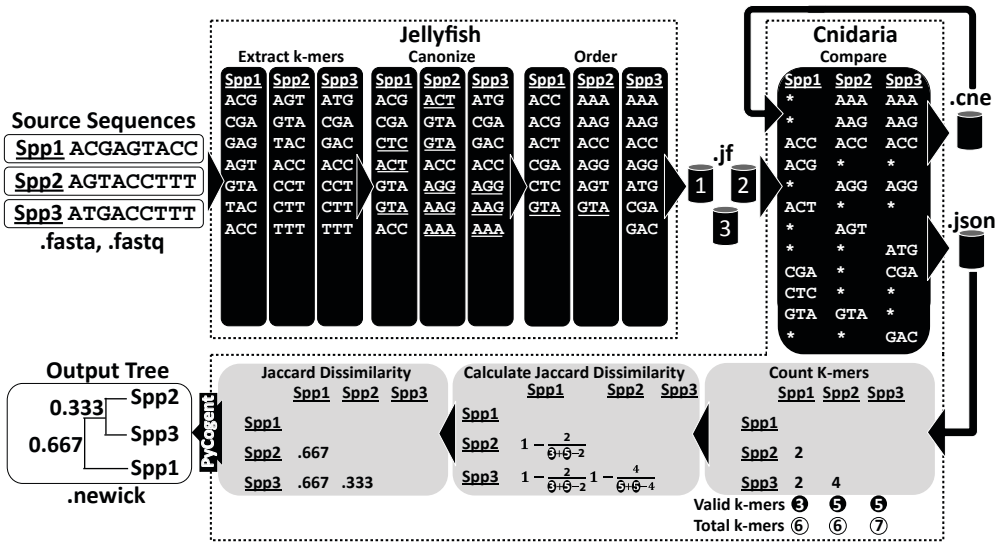

**Figure 1:** Cnidaria analysis summary. The Jellyfish software reads each of the source sequence files (in Fasta or Fastq formats), extracts their *k*-mers (*k* = 3 in this example), canonizes them (by generating the reverse complement of each *k*-mer and storing only the *k*-mer which appears first lexicographically), orders them according to a deterministic hashing algorithm (in this example, alphabetically) and then saves each dataset in a separated database file (.jf). Cnidaria subsequently reads these databases and compares them, side-by-side, by counting the total number of *k*-mers (white circles), the number of valid *k*-mers (*k*-mers shared by at least two samples, black circles) and the number of shared *k*-mers for each pair of samples as a matrix. Those values are exported to a Cnidaria Summary Database (CSD, a .json file) that is then used to construct a matrix of Jaccard distances between the samples (Formula 1). This dissimilarity matrix is then used for clustering by Neighbour-Joining and exported as a Newick tree. Alternatively, Cnidaria can export a Cnidaria Complete Database (CCD, a .cne file) containing all *k*-mers and a linked list describing their presence/absence in the samples. This second database can be used as an input dataset together with other .cne or .jf files.

Besides the source code, a precompiled version of Cnidaria is available that runs on most 64-bit Linux distributions. To use this, NGS and genome files should first be converted to a JELLYFISH database using JELLYFISH (a precompiled version 2.13 is also included with CNIDARIA, along with auxiliary scripts to facilitate the conversion). Then, CNIDARIA should be run on as many CCD databases as needed, either in Database Creation Mode (producing both a CSD and CCD file) or in Sample Analysis Mode (producing only a CSD file). Either way, an auxiliary python script can be run on the CSD file to generate a $k$-mer count matrix in CSV format, a *Jaccard* distance matrix in CSV format and a phylogenetic tree in NEWICK format. Helper scripts are available to visualize the NEWICK tree as a PNG file.

## Results and Discussion
### Data set

To validate the performance of CNIDARIA, we gathered a collection of 135 genomic, transcriptomic and raw NGS datasets covering a wide range of organisms. 84 samples were of tomatoes, part of a large dataset recently published (Aflitos *et al.*, 2014). A list of all samples can be found in Supplementary Table 1. All datasets were analysed using JELLYFISH counting canonized $k$-mers. Canonization is the process of only storing the lexicographically smallest between a $k$-mer and its reverse complement. This step is required as both molecules are technically the same: the existence of one implies the existence of the other on the complementary DNA strand. The datasets were then split in 50 pieces and divided over 20 threads on an 80 core Intel(R) Xeon(R) CPU E7- 4850 @ 2.00 GHz machine, speeding up the analysis approximately 40 times compared to single-thread analysis on the same CPU. We then created a CNIDARIA Complete Database (CCD) containing all 135 samples.

### Influence of *k*-mer size

To investigate the influence of the $k$-mer size in the accuracy of the phylogenetic inference of CNIDARIA, we analysed the panel of 135 samples with $k$ = 11, 15, 17, 21 and 31 (predefined hash sizes of 128 million, 256 million, 512 million, 1 billion and 4 billion, respectively). The resulting statistics can also be found in Supplementary Table 1.

Due to the low complexity of 11-mers, all possible $k$-mer of this size were found in the datasets and all $k$-mers were shared between at least two samples (Table 1). This carries little clustering information and generates many zero distances (minimum dissimilarity) as shown in Figures 2, 3 and 4. Phylogenetic distances increase with $k$-mer size and 31-mers have most distances equal to 1, i.e. maximum dissimilarity (except for highly related species), which does not allow clustering of distant species.

### Identification accuracy

To classify samples, we used the 1-nearest neighbour algorithm on 30 samples for supraspecies level analysis (8 genus, 7 families, 7 orders, 4 phylum and 3 kingdoms, summa-rized in Table 2) and on 33 samples for species level analysis (11 species of the Solanum clade, described in Supplementary Tables 2 and 3, summarized in Table 2). The 1-nearest neighbour classifier assigns to each sample, for each phylogenetic level (species, genus, family, order, phylum and kingdom), the phylogenetic class of the sample with the smallest distance. We report the percentage of samples correctly classified. 15-mers and 17-mers yielded accuracy above 70% and 90%, respectively, at the supra-species level but accuracy below 75% at the species level. Both 21- and 31-mers allowed us to correctly classify 100% of the samples at the supra-species level and 78% of the samples at the species level (Figure 5). The lower accuracy for species level classification in the tomato clade can be attributed to introgressions and sympatric speciation in tomato, which is reflected in the clustering distance data (Figure 6) and is in agreement with the clustering obtained by (Aflitos *et al.*, 2014) which used whole genome SNP analysis to construct trees. The 31-mers dataset, when compared to the 21-mers dataset, resulted in an increased run time and disk usage without giving a discernible higher discriminative power (Figures 2 ,3 and 4). This suggests 21 is a good *k*-mer size for general purpose clustering. Conversely, 31-mers are frequently used for NGS data quality checking (reviewed in Leggett *et al.*, 2013) and the same Jellyfish database can be used for species identification.

**Table 1:** Summary of search space per $k$-mer size and number of $k$-mers found in datasets. The second column contains the total number of possible $k$-mers, calculated as ($4^{k\text{-mer size}}$ / 2), where the division by two is due to canonization. The third column is the median and the Median Absolute Deviation (MAD) of the total number of $k$-mers found in the samples divided by the number of possible $k$-mers (second column), showing the percentage of combinations actually found and, consequently, the saturation of the search space; the fourth column gives the median and MAD of the percentage of valid $k$-mers (shared between at least two samples).

| $k$-mer size | # Canonical $k$-mer combinations | % of $k$-mers found per sample | | % of $k$-mers shared by at least two samples | |
|---|---|---|---|---|---|
| | | **Median** | **MAD** | **Median** | **MAD** |
| **11-mer** | $2.1 \times 10^{06}$ | 100.00% | 1.58% | 100.00% | 0.00% |
| **15-mer** | $5.4 \times 10^{08}$ | 53.59% | 17.07% | 100.00% | 0.00% |
| **17-mer** | $8.6 \times 10^{09}$ | 8.90% | 4.03% | 98.37% | 0.99% |
| **21-mer** | $2.2 \times 10^{12}$ | 0.05% | 0.03% | 81.45% | 20.55% |
| **31-mer** | $2.3 \times 10^{18}$ | 0.000000061% | 0.000000032% | 67.05% | 24.14% |

**Table 2:** Percentage of correct classification for each k-mer size and each taxonomic level. Sample names are found in Supplementary Tables 2 and 3.

| K-mer size | Species | Genus | Family | Order | Phylum | Kingdom |
|---|---|---|---|---|---|---|
| **11-mer** | 0% | 57% | 63% | 63% | 77% | 77% |
| **15-mer** | 55% | 73% | 80% | 80% | 80% | 80% |
| **17-mer** | 73% | 93% | 97% | 97% | 97% | 97% |
| **21-mer** | 79% | 100% | 100% | 100% | 100% | 100% |
| **31-mer** | 79% | 100% | 100% | 100% | 100% | 100% |

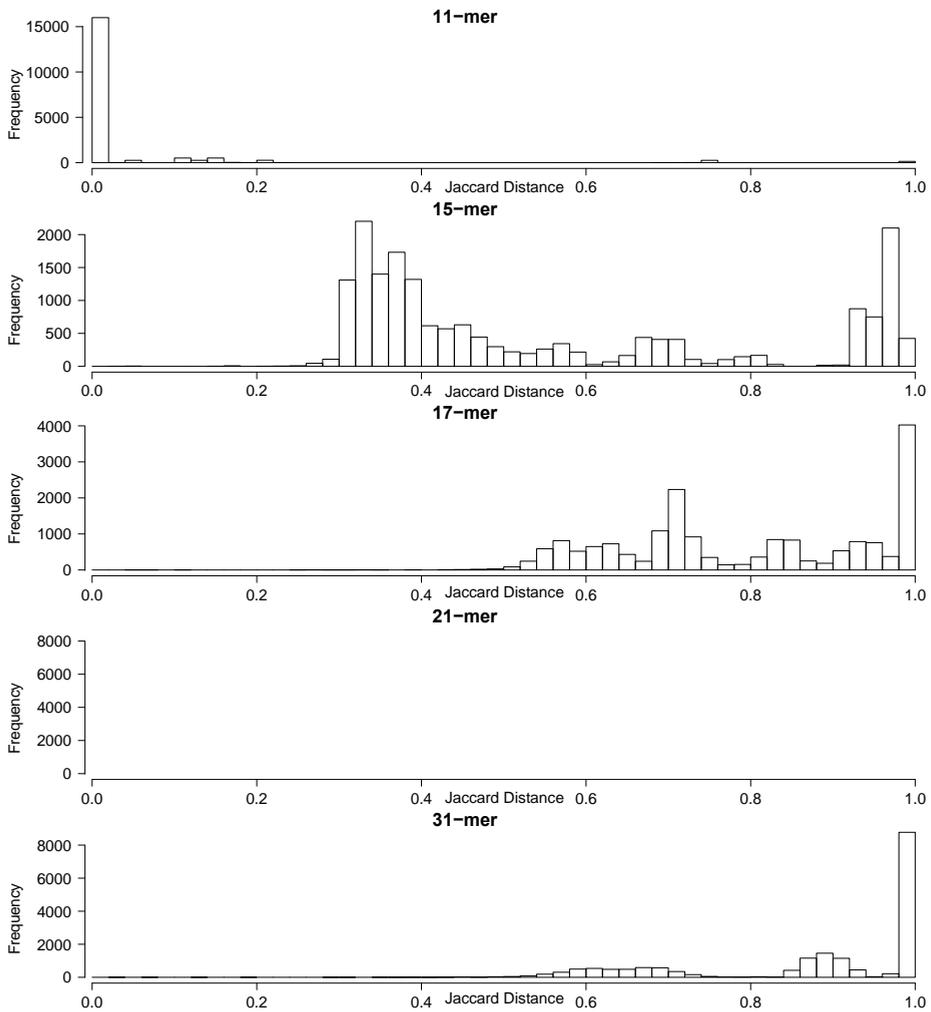

**Figure 2:** Histogram of Jaccard distances for each k-mer size of the 135 samples. A distance of 0 means identity while a distance of 1 means no similarity. Using 11-mers most samples are identical to each other. For 31-mers, most samples share no similarity with any other sample except for phylogenetically closely related samples. 17 and 21-mers show higher similarity between groups.

**3A**

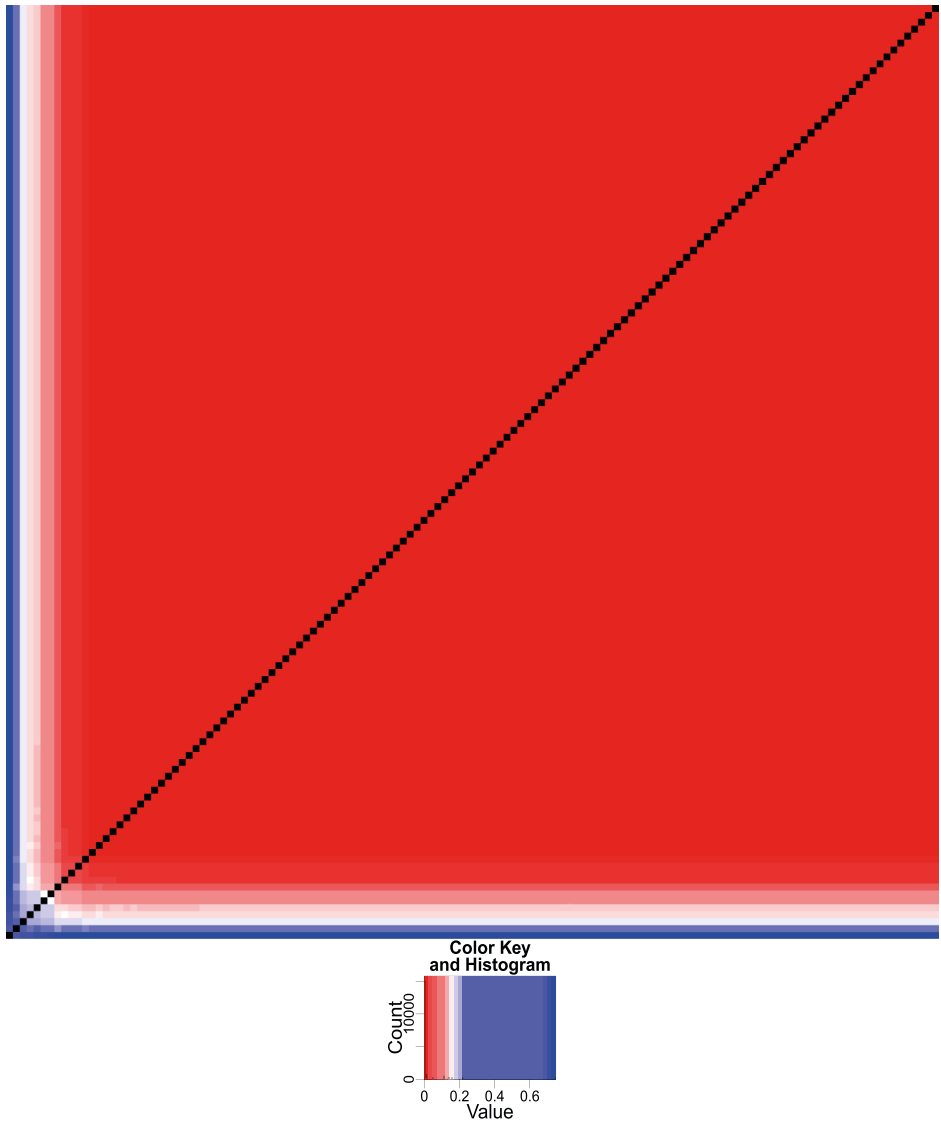

**Figure 3:** Heatmaps of Jaccard distance of 135 samples using 11-, 15-, 17-, 21- and 31-mers are shown in graphs A, B, C, D and E, respectively. Here, 0 (red) means identity between samples while 1 (blue) means no identity. Generally, closely related species show high similarity with closely related species and no similarity with outgroups. This leads to strong clustering inside groups but loose coupling between groups.

**3B**

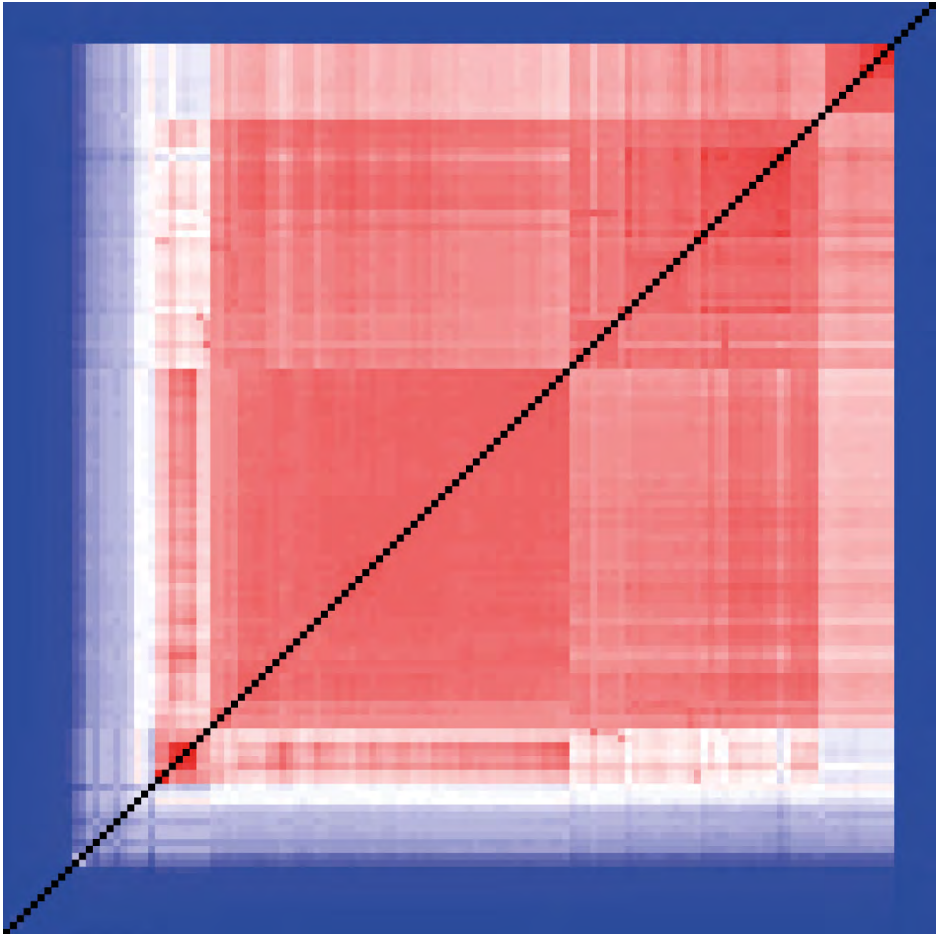

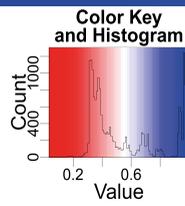

**3C**

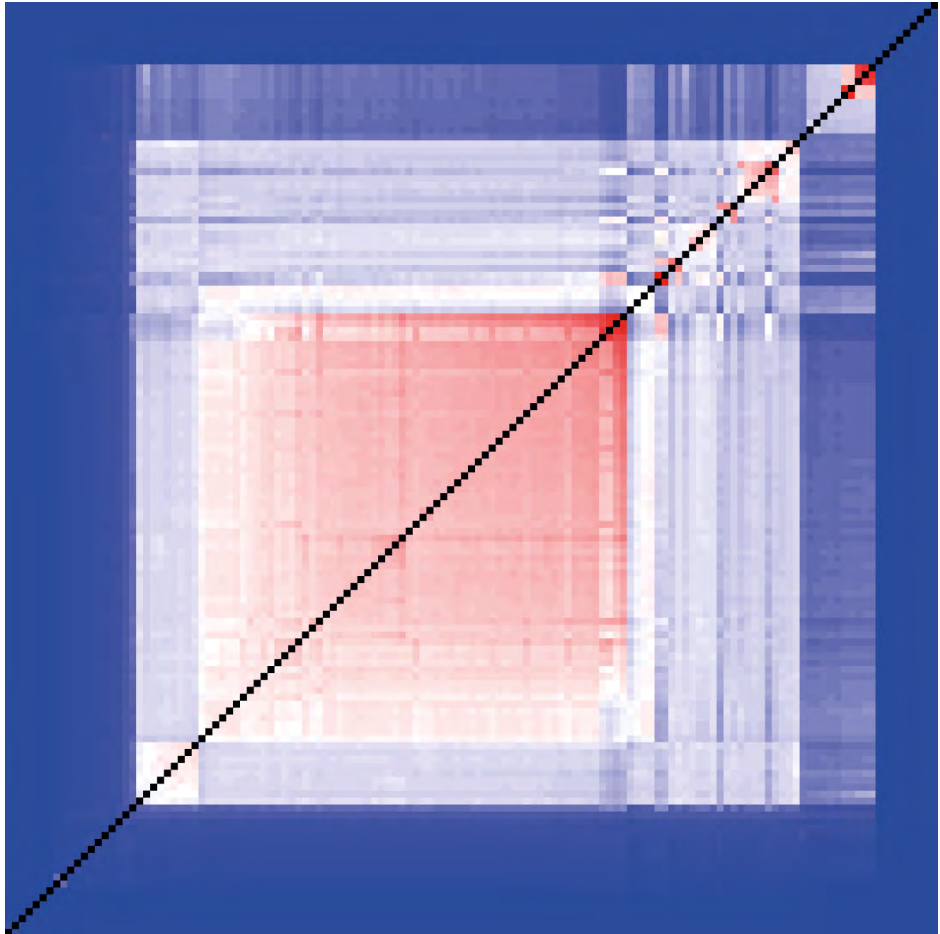

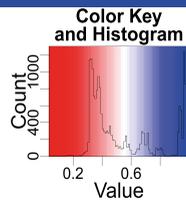

**3D**

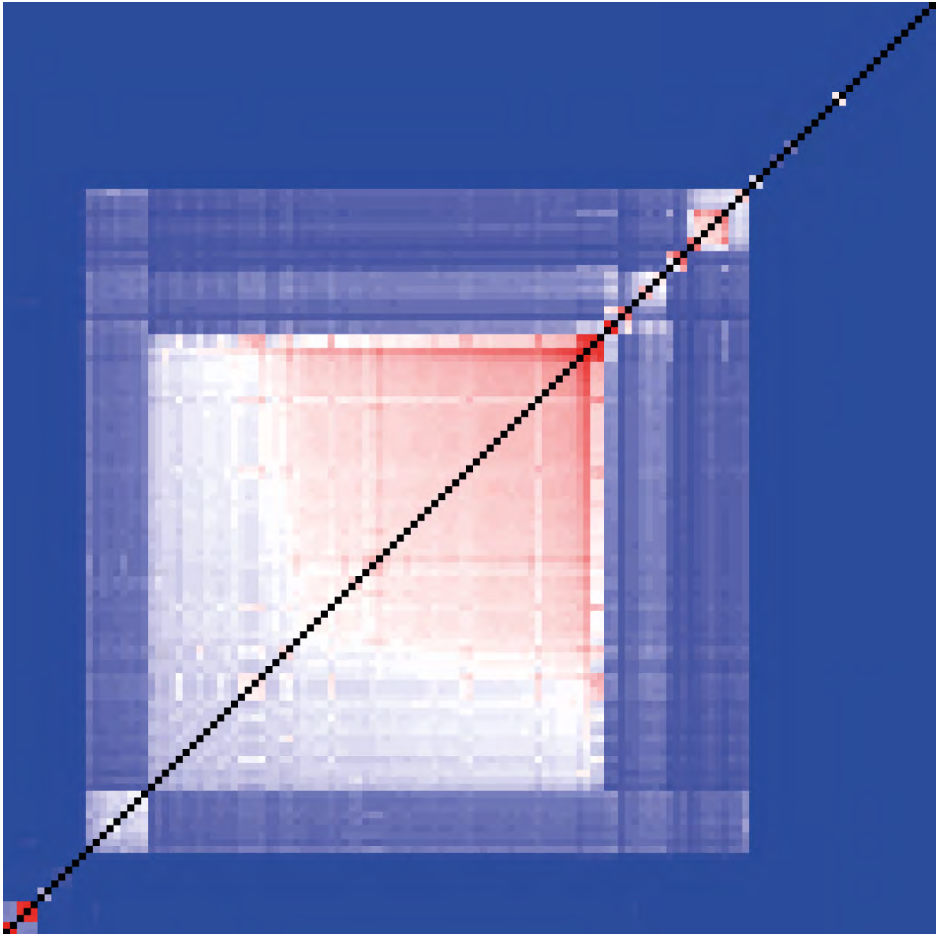

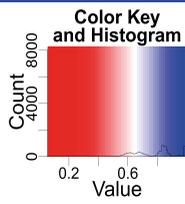

**3E**

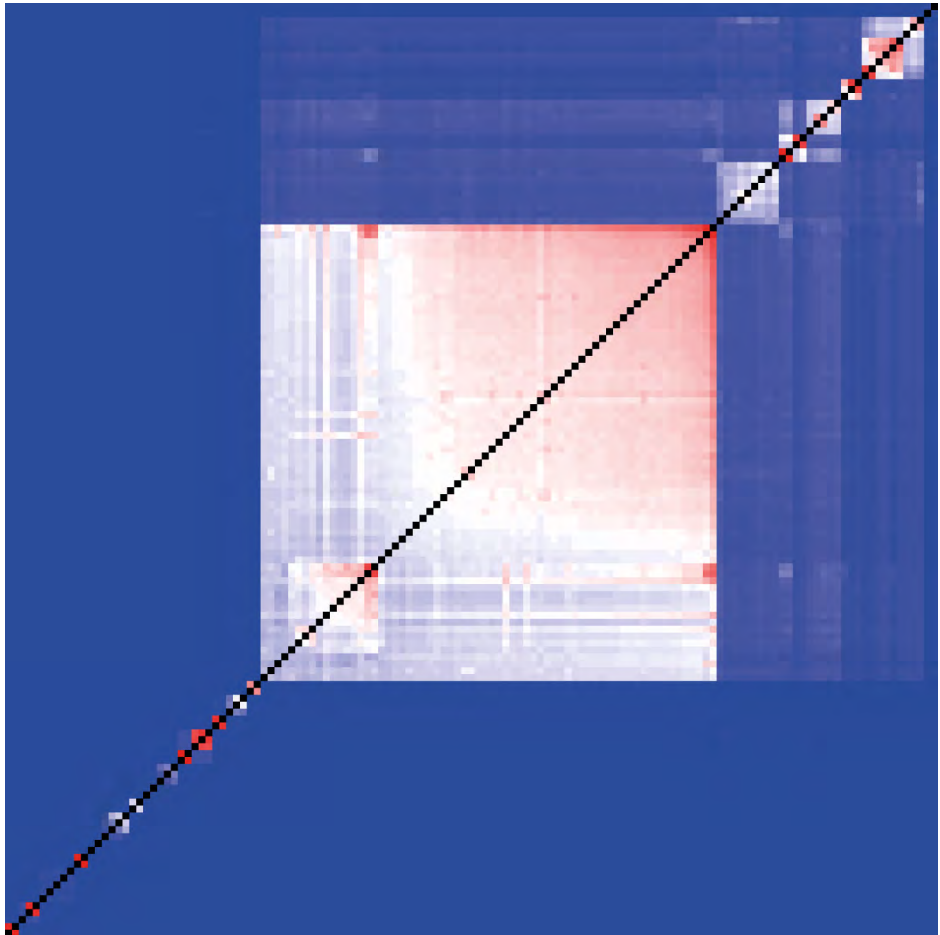

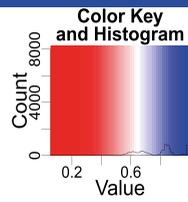

## 4.A.I

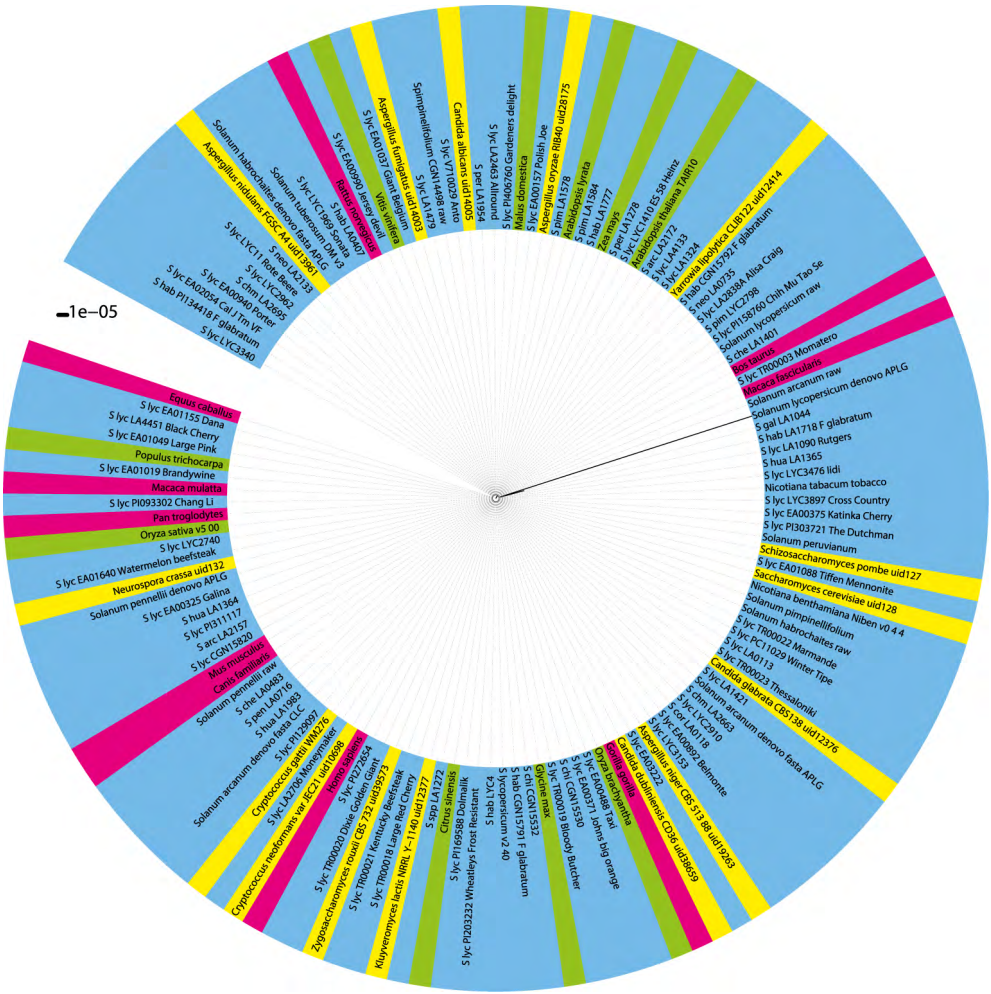

**Figure 4:** Phylogenetic trees from 135 samples. 11-, 15-, 17-, 21- and 31-mers are shown in graphs A, B, C, D and E, respectively. I) tree with phylogenetic distances; II) tree without phylogenetic distances, showing the clustering more clearly; trees plotted using iTOL (Letunic and Bork, 2007).

Solanaceae
Non-solanaceae Plant
Fungi
Animal

## 4.B.I

Solanaceae

Non–solanaceae Plant

Fungi

Animal

## 4.C.I

Solanaceae
Non−solanaceae Plant
Fungi
Animal

## 4.D.I

**4.D.II**

Solanaceae

Non−solanaceae Plant

Fungi

Animal

## 4.E.I

Solanaceae

Non−solanaceae Plant

Fungi

Animal

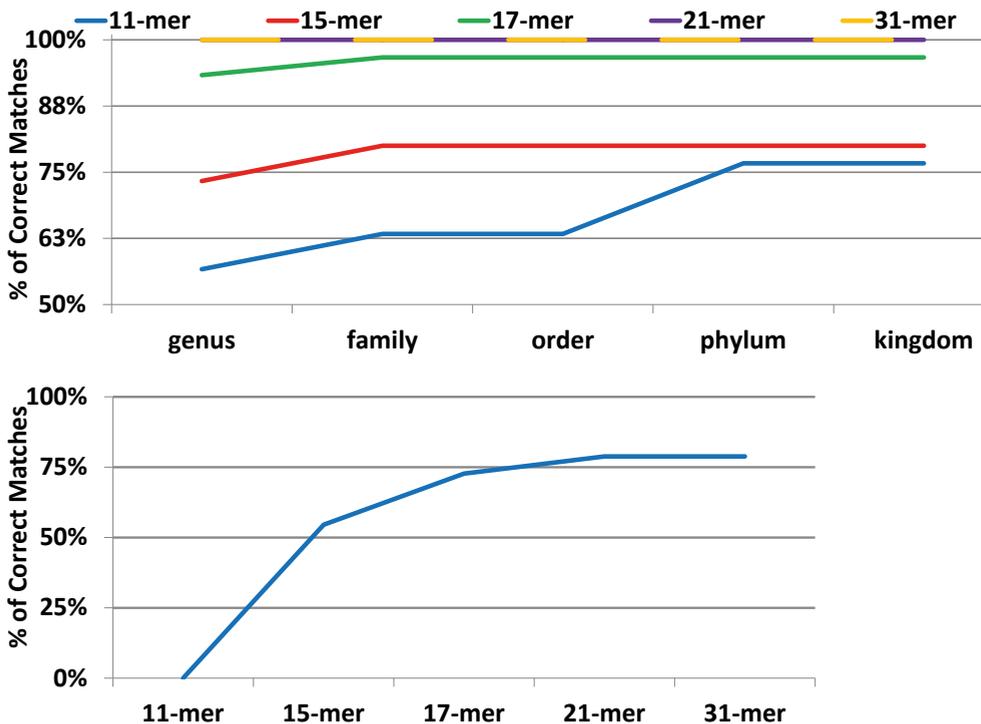

**Figure 5:** 1-nearest-neighbour analysis for the supra-species level and species level. Top shows supra-species level analysis of 30 samples, 8 genus, 7 families, 7 orders, 4 phylum and 3 kingdoms. Bottom shows species level analysis of 33 samples from 11 species of the Solanum clade. Classification reports the Leave-One-Out Cross-Validation error estimate (LOOCV), showing the improvement in accuracy with increased k-mer size and flattening of precision for k-mers above 21. Sample names and classes can be found in supplementary tables 3 and 4, respectively.

## Speedup by subsampling

To test the influence of data set size (and possibility of speedup) and since 21-mers showed the best trade-off between speed and discriminating power (consistent with Cannon *et al.*, 2010), we sampled 2% of the dataset by analysing just 1 of the 50 pieces the data was originally split into. Figure 7 shows the phylogenetic placement of species in the trees constructed using this dataset. The tree is indistinguishable from the one generated on the full dataset, illustrating the ability of CNIDARIA to correctly classify samples even at very low sequencing coverage. This suggests that CNIDARIA should be able to correctly cluster and identify samples using small and affordable NGS sequencing technology such as Illumina MiSeq nano runs (500 Mbp in 2x250bp reads, Illumina, 2015).

## Joint analysis of DNA and rna-seq data

Next, we expanded the 135 sample dataset (built using Database Creation Mode) with 34 extra samples, 26 genomic and 8 RNA-seq (Supplementary Table 1), using 21-mers and the faster Sample Analysis Mode. RNA-seq samples were added to verify whether transcriptome data would cluster with their genomic NGS counterparts, despite their small coverage of the genome length. Results are shown in Figure 8. The clustering of the original 135 samples is not changed and new samples cluster correctly according to their phylogeny. The consistent clustering observed for the RNA-seq dataset illustrates the ability of CNIDARIA to use such data for accurate species identification.

## Comparison with REFERENCEFREE

To demonstrate the advantages of CNIDARIA, we compare it to a state-of-the-art alternative tool, proposed in (Cannon *et al.*, 2010). The latest version of the software introduced (version 1.1.3, hereafter referred to simply as REFERENCEFREE) was downloaded and run in conjunction with ABYSS (Simpson *et al.*, 2009). We used ABYSS version 1.3.3 rather than the latest version (1.9.0) since that was the last version tested with REFERENCEFREE. REFERENCEFREE was run single threaded on an Intel(R) Xeon(R) CPU E7- 4850 @ 2.00 GHz with a $k$-mer size of 21, a minimum frequency of 0 (i.e. using all $k$-mers appearing 1 or more times), no complexity filter and no sampling of $k$-mers. The list of shared $k$-mers generated was then parsed using the CNIDARIA scripts in order to generate the phylogenetic tree.

Using a subset of our data (41 assembled genomes, Supplementary Table 1) containing 40 Gbp and 20 billion $k$-mers, REFERENCEFREE (Supplementary Table 1) and JELLYFISH have a comparable speed for $k$-mer counting, taking 4 hours to count 445 million $k$-mers (2% of the total). REFERENCEFREE then took 60% more time than CNIDARIA in single threaded Sample Analysis Mode for merging and summarizing the results (70 hours vs. 44 hours, respectively). Note that the databases created by CNIDARIA can be re-used in subsequent comparisons, whereas REFERENCEFREE requires all the $k$-mer count files to be merged again when re-run. Moreover, CNIDARIA has the important advantage of being highly parallelizable while REFERENCEFREE can only be run single threaded. The phylogenetic tree created by REFERENCEFREE can be found in figure 9.

Besides speed, CNIDARIA (and JELLYFISH) use significantly less space, due to their binary formats. They generate files which are smaller than the equivalent files created by REFERENCEFREE with median sizes of 9.2 Gb vs. 42.2 Gb (MADs of 2.5 Gb and 11.0 Gb, respectively) for the $k$-mer count file and 227 Gb vs. 2.1 Tb for the merged $k$-mer count file, despite the merged $k$-mer count file created by REFERENCEFREE contain only 2% of the total number of $k$-mers.

**6A**

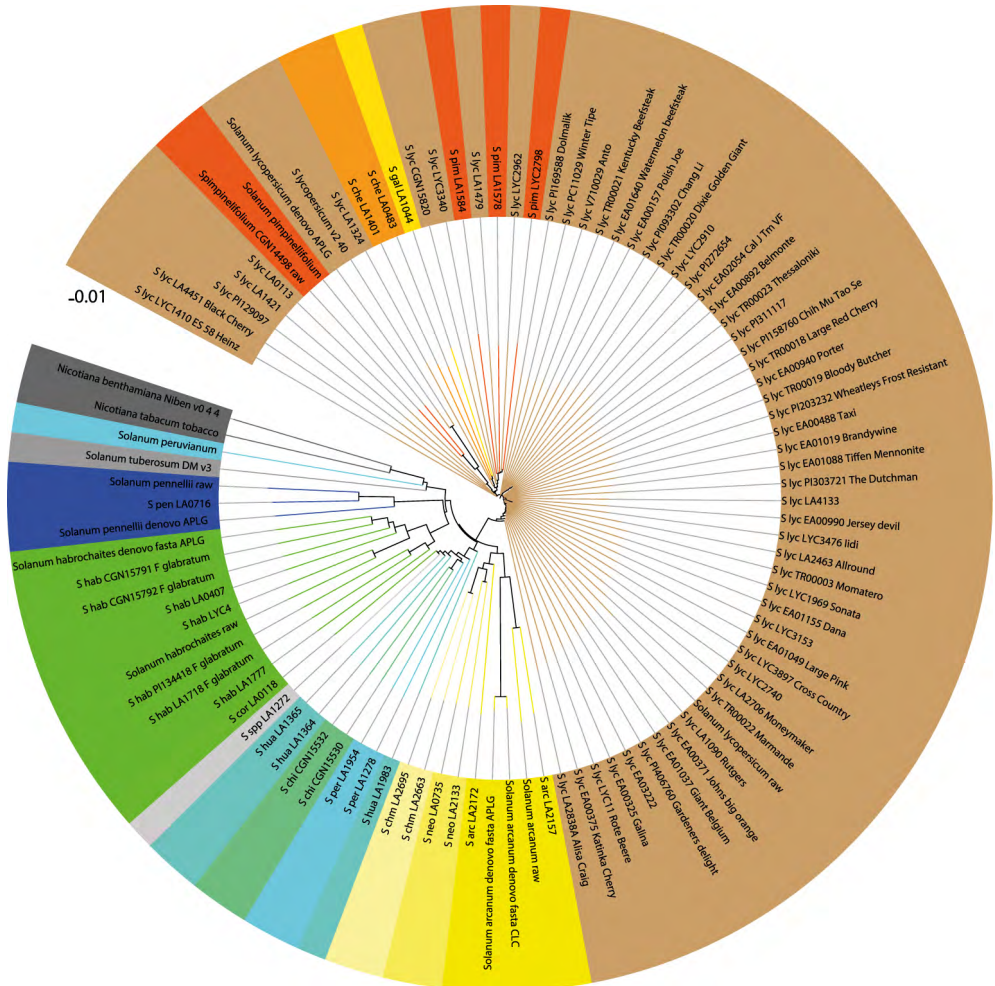

**Figure 6:** Phylogenetic tress with and without branch lengths of 98 Solanum taxa from 13 species. The Lycopersicon group (comprised of *Solanum lycopersicum*, *S. pimpinellifolium*, *S. cheesmaniae* and *S. galapagense*) clusters as a monophyletic group. Sometimes the non-*S. lycopersicum* species cluster inside the *S. lycopersicum* clade. We speculate these are *S. lycopersicum* varieties containing intro-gression clustering with the donor species, consistently with the findings of Aflitos *et al.* (2014). The Arcanum group (comprised of *S. arcanum*, *S. chmielewskii* and *S. neorikii*) also clusters monophyleti-cally, closer to the Eriopersicon group, its sister group. The North Eriopersicon group (comprised of *S. huaylasense*, *S. chilense*, *S. peruvianum* and *S. corneliomulleri*) groups with the South Eriopersicon group (comprised of *S. habrochaites*, its only member) and its sister group, Neolycopersicon (comprised of *S. pennelli*, its only member). *S. tuberosum* and *S. nicotiana* were added as outgroups. Sample names ending in RAW are raw genomic data; names ending in APLG and CLC are assembled genomes. A) tree with phylogenetic distances; B) tree withtout phylogenetic distances, showing the clustering more clearly; Trees were plotted using iTOL (Letunic and Bork, 2007).

**6B**

Solanum lycopersicum

Solanum neorickii

Solanum chmielewskii

Solanum arcanum

Solanum huaylasense

Solanum chilense

Solanum peruvianum

Solanum spp

Solanum corneliomuelleri

Solanum habrochaites

Solanum pennellii

Solanum tuberosum

Nicotianeae

Solanum cheesmaniae

Solanum galapagense

Solanum pimpinellifolium

**7A**

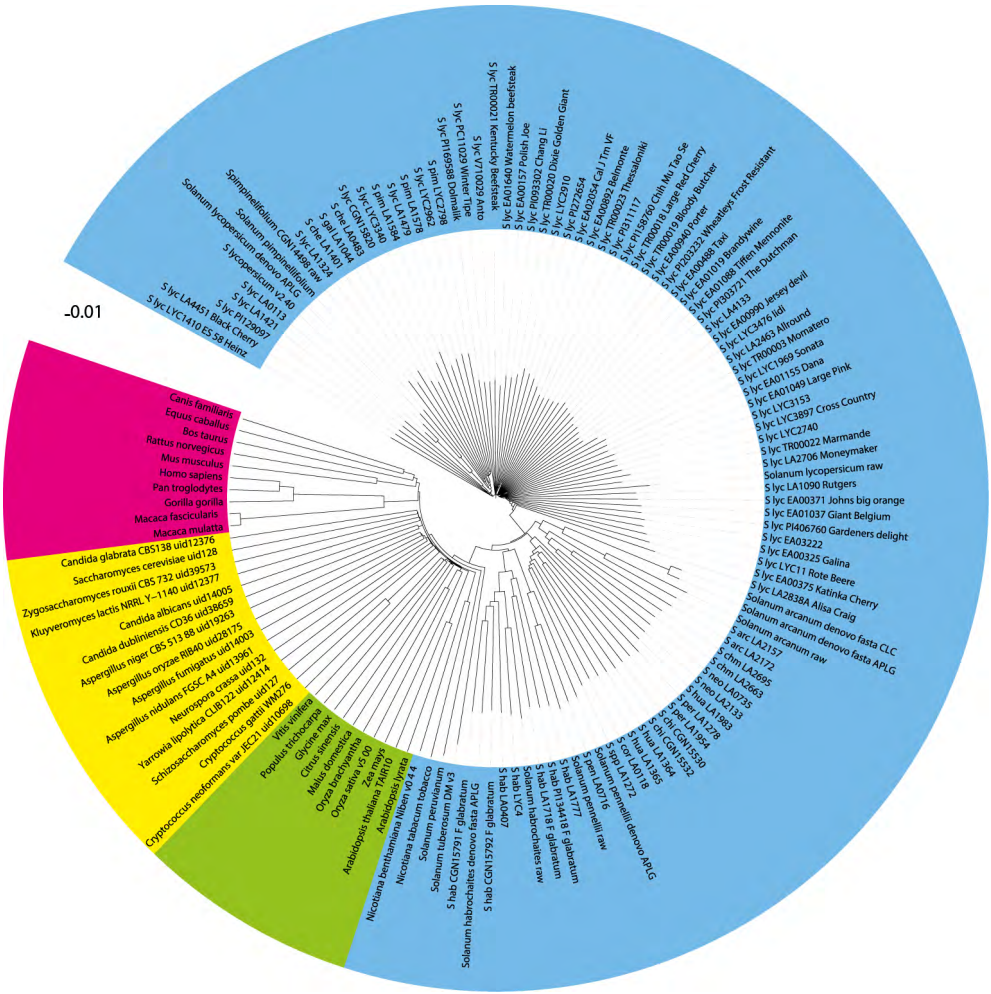

**Figure 7:** Results for 2% of the 21-mer dataset. A) phylogenetic tree with distance; B) phylogenetic tree without distance; C) heatmap of phylogenetic distances showing low inter-group similarity and high intra-group similarity; D) histogram of Jaccard distances showing the same feature of low inter-group similarity and high intra-group similarity. Trees were plotted using iTOL (Letunic and Bork, 2007).

Solanaceae
Non−Solanaceae Plant
Fungi
Animal

**7C**

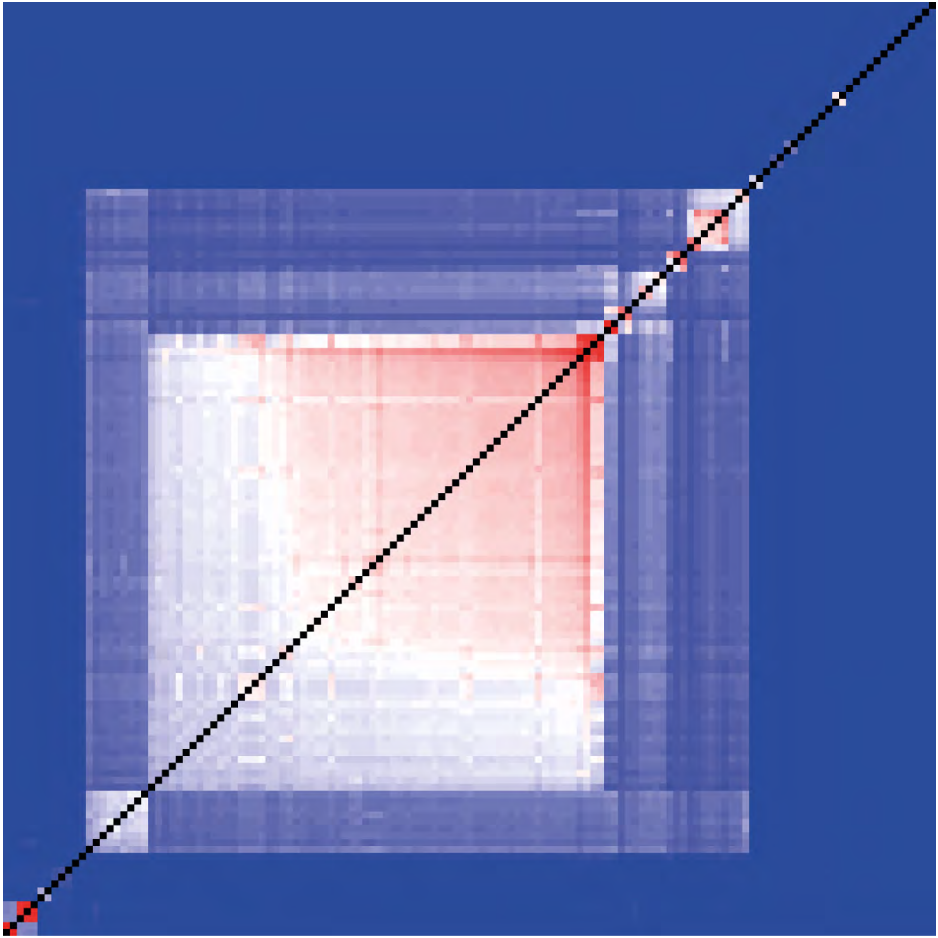

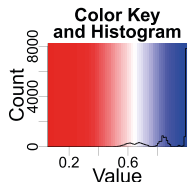

**7D**

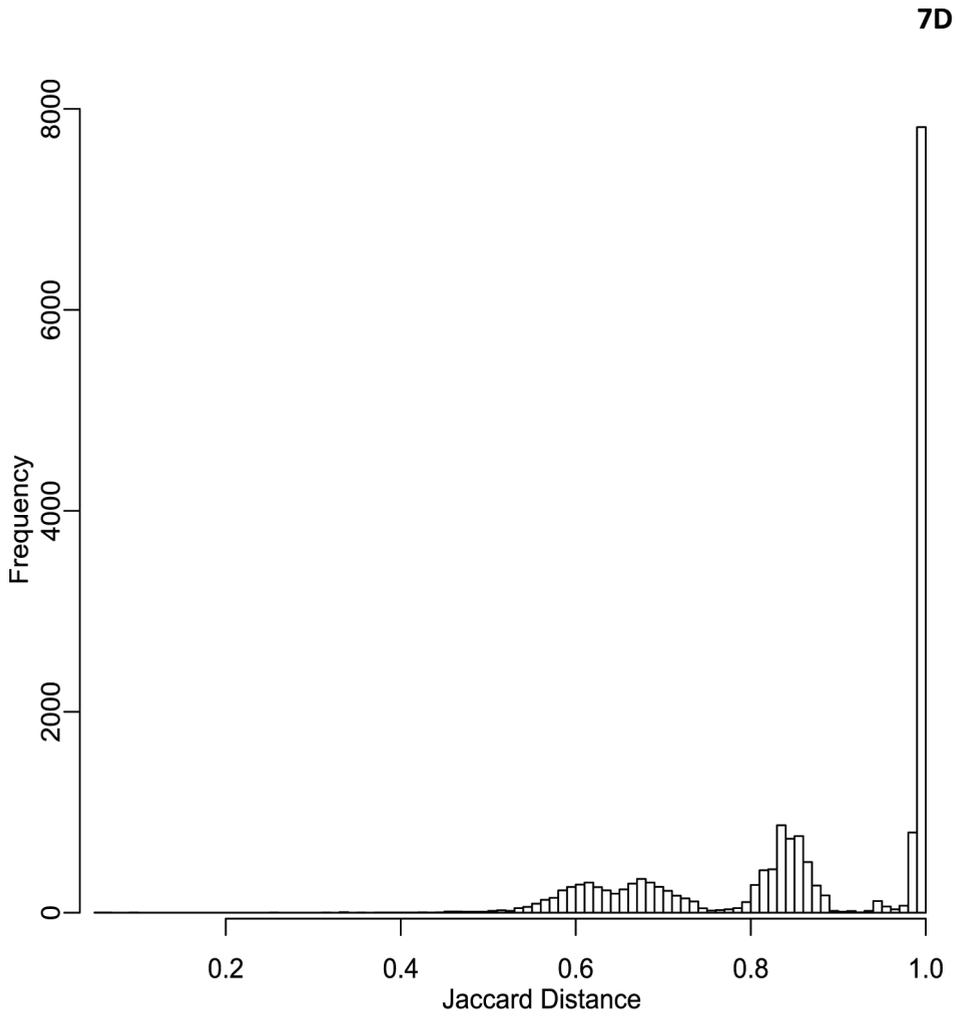

**8A**

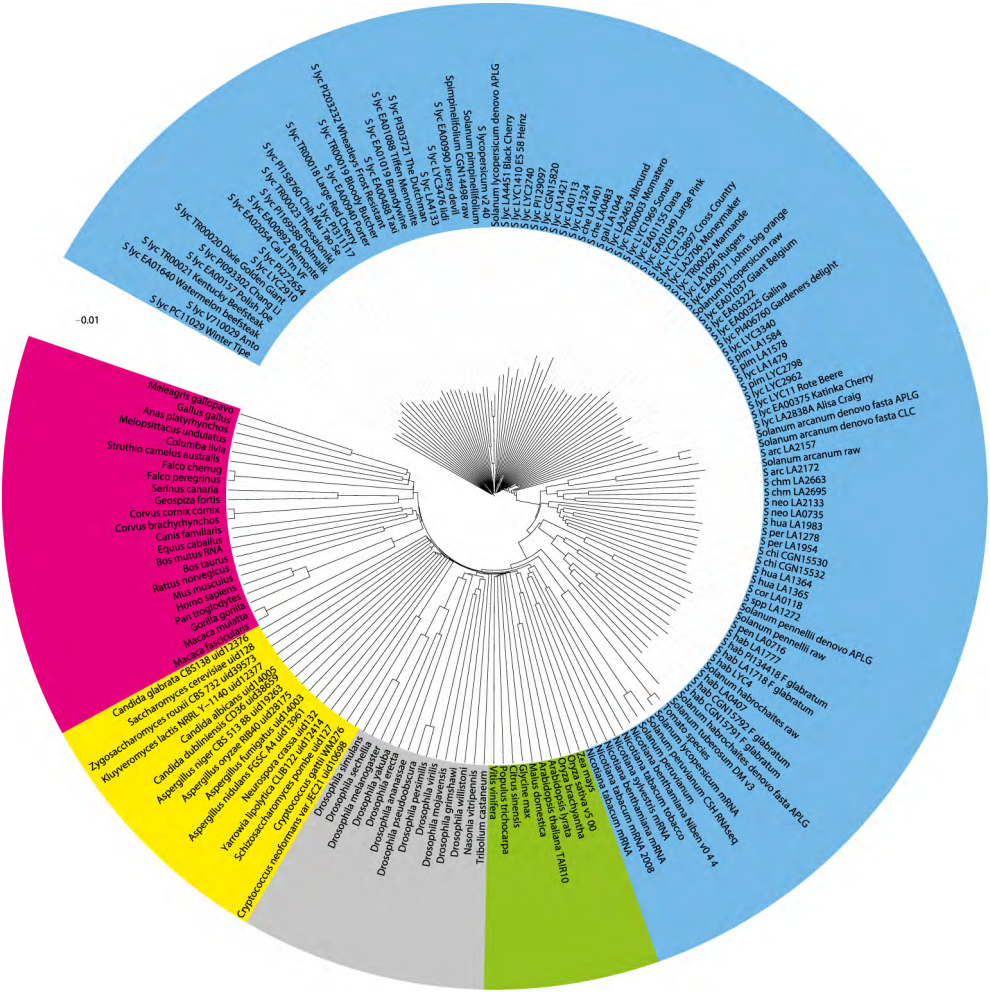

**Figure 8:** Results for the 21-mer dataset of 169 individuals using Jaccard distance and Neighbour-Joining. A) phylogenetic tree with distance; B) phylogenetic tree without distance (tree branch length); C) heatmap of phylogenetic distances showing low inter-group similarity and high intra-group similarity; D) histogram of Jaccard distances showing the same feature of low inter-group similarity and high intra-group similarity. Sample names ending in RAW are raw genomic data; names ending in APLG and CLC are assembled genomes; names ending in RNA, RNAseq and mRNA are RNA-seq datasets. Trees were plotted using iTOL (Letunic and Bork, 2007).

**8B**

Solanaceae
Fungi
Non−Solanaceae Plant
Insect
Animal

**8C**

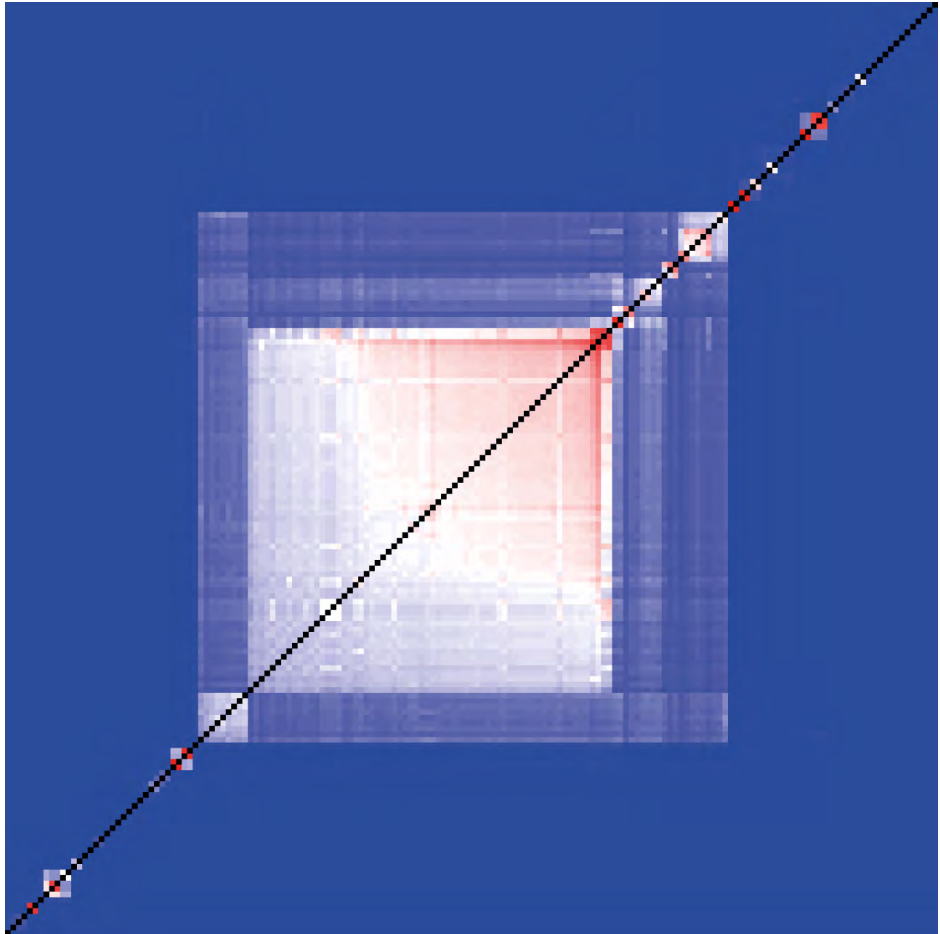

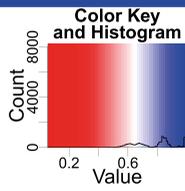

**8D**

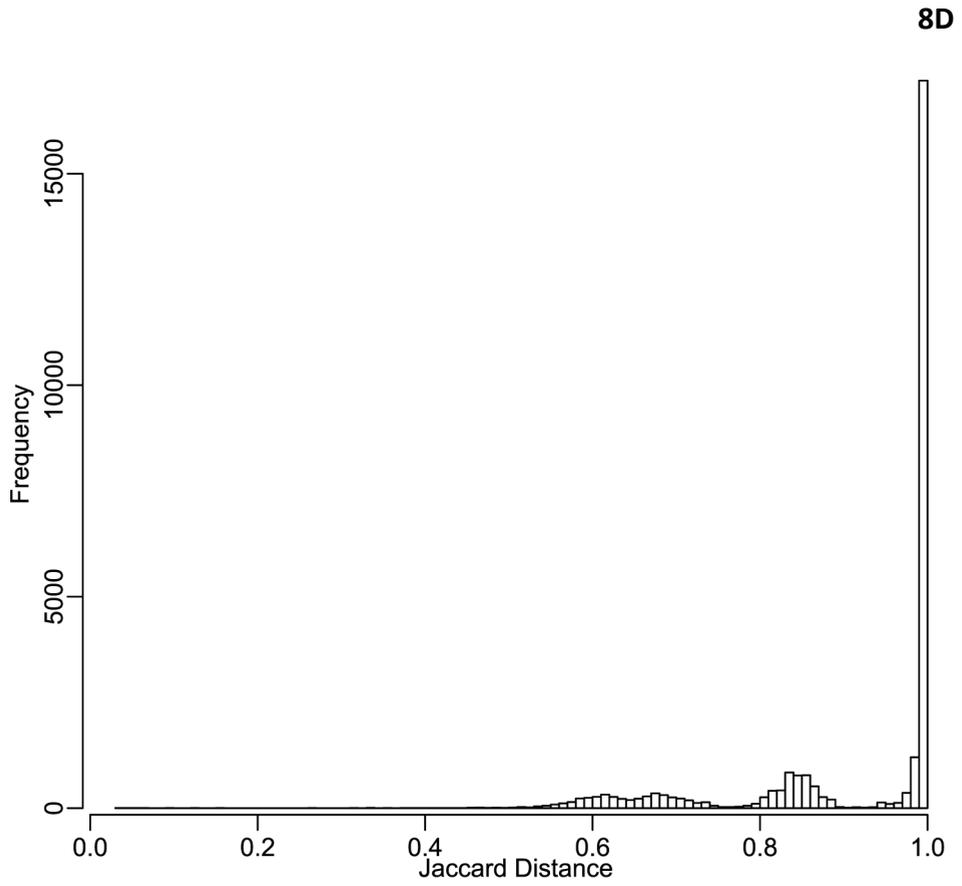

**Figure 9:** Phylogenetic tree created using REFERENCEFREE for 41 genomes. Due to filtering and sampling performed by REFERENCEFREE, clustering is impaired and inconsistent as exemplified by the positions of *Pan troglodites*, Nicotiana and Aspergillus.

## Conclusion

We have introduced Cnidaria, a tool to quickly and reliably analyse WGS and RNA-seq samples from both assembled and unassembled NGS data, offering significant advantages in terms of time and space requirements compared to a state-of-the-art tool. By clustering in total 169 eukaryotic samples from 78 species (42 genus, 32 families, 27 orders, 5 phyla, 6 divisions and 3 kingdoms from the Eukaryota superkingdom) we have demonstrated that Cnidaria can handle a large number of samples from very distant phylogenetic origins, producing a reliable tree with up to 100% classification accuracy at the supra species level and 78% accuracy at the species level, the later value being low mostly due to interspecific crossings. As Cnidaria is also able to analyse RNA-seq data, researchers can acquire, besides the species information, physiological state information such as pathogenicity and stress response of the sample for downstream analysis.

A database created in Database Creation Mode allows querying directly for *k*-mers shared by a specified set of samples, enabling comparisons useful in several applications. Examples include identifying and quantifying polymorphisms between closely related samples, quantifying sequence diversity in the setup phase of large sequencing projects for sample selection, and ecological diversity analysis. In addition, *k*-mers shared exclusively by a set of samples can be used for diagnostic primer design, supporting the detection of target genes. Furthermore, mismatching *k*-mers between a sample and a close relative can be used to identify the source of contamination or introgressions, as performed by (Byrd *et al.*, 2014).

### Availability and requirements

| | |
|---|---|
| Project name: | Cnidaria |
| Project home page: | http://www.ab.wur.nl/cnidaria |
| Operating system(s): | 64-bit Linux |
| Programming language: | C++ x11 and Python 2.7 |
| Other requirements: | None to run; GCC 4.8 or higher for compiling |
| License: | MIT |
| Restrictions to use by non-academics: | No |

### List of Abbreviations Used

| | |
|---|---|
| CCD | = Cnidaria Complete Database |
| CSD | = Cnidaria Summary Database |
| CSV | = Comma-Separated Values |
| DFT | = Discrete Fourier Transform |
| EST | = Expressed Sequence Tag |
| *K*-mer | = Substring of size k |
| MAD | = Median Absolute Deviation |
| NGS | = Next Generation Sequencing |
| PNG | = Portable Network Graphics |
| rDNA | = ribosomal DNA |
| RFM | = Reference-Free Methods |
| RNA-seq | = RNA sequencing |
| WGS | = Whole Genome Sequencing |
| .fastq | = Raw assembly file |
| .fasta | = Assembled sequence |
| .Newick | = Phylogenetic tree file format |
| HTML | = HyperText Markup Language |
| Read | = Contiguous sequence outputted by sequencing machine |

## Competing interests

The author(s) declare that they have no competing interests

## Authors' contributions

SAA designed, wrote and tested the software; DR participated in the design of the software and validated the algorithm; ES, GSP, SP and HJ participated in the design of the software; All authors contributed in the confection of the manuscript, read it and approved the final manuscript.

## Authors' information

**Saulo Alves Aflitos**  PhD candidate in Bioinformatics

**Edouard Severing**  PhD in Bioinformatics, Post-doc in Max-Planck-Institut für Pflanzenzüchtungsforschung - Köln

**Gabino Sanchez-Perez**  PhD in Bioinformatics - Senior Researcher Bioinformatics at Wageningen University, Cluster leader Cluster Bioinformatics – Plant Research International – Wageningen University

**Sander Peters**  PhD in Bioinformatics - Senior scientist / Bioinformatician at Plant Research International – Wageningen University

**Hans de Jong**  Professor of Cytogenetics at Wageningen University

**Dick de Ridder**  Professor of Bioinformatics at Wageningen University

## Acknowledgements

This project was funded by Centre for BioSystems Genomics (CBSG) under the grant number TO09.